# Nanofluidic Platform for Studying the First-Order Phase Transitions in Superfluid Helium-3

Petri J. Heikkinen[1] · Nathan Eng[1] · Lev V. Levitin[1] · Xavier Rojas[1] · Angadjit Singh[1,2] · Samuli Autti[3] · Richard P. Haley[3] · Mark Hindmarsh[4,5] · Dmitry E. Zmeev[3] · Jeevak M. Parpia[6] · Andrew Casey[1] · John Saunders[1]

## Abstract

The symmetry-breaking first-order phase transition between superfluid phases $^3$He-A and $^3$He-B can be triggered extrinsically by ionising radiation or heterogeneous nucleation arising from the details of the sample cell construction. However, the role of potential homogeneous intrinsic nucleation mechanisms remains elusive. Discovering and resolving the intrinsic processes may have cosmological consequences, since an analogous first-order phase transition, and the production of gravitational waves, has been predicted for the very early stages of the expanding Universe in many extensions of the Standard Model of particle physics. Here we introduce a new approach for probing the phase transition in superfluid $^3$He. The setup consists of a novel stepped-height nanofluidic sample container with close to atomically smooth walls. The $^3$He is confined in five tiny nanofabricated volumes and assayed non-invasively by NMR. Tuning of the state of $^3$He by confinement is used to isolate each of these five volumes so that the phase transitions in them can occur independently and free from any obvious sources of heterogeneous nucleation. The small volumes also ensure that the transitions triggered by ionising radiation are strongly suppressed. Here we present the preliminary measurements using this setup, showing both strong supercooling of $^3$He-A and superheating of $^3$He-B, with stochastic processes dominating the phase transitions between the two. The objective is to study the nucleation as a function of temperature and pressure over the full phase diagram, to both better test the proposed extrinsic mechanisms and seek potential parallel intrinsic mechanisms.

**Keywords** Helium-3 · First-order phase transition · Nanofabrication · Confinement

Extended author information available on the last page of the article

## 1 Introduction

The mechanism behind the first-order phase transition between the superfluid A and B phases of $^3$He has evaded explanation despite decades of both experimental and theoretical work. Moreover, cosmological analogues [1–5] connect this fundamental problem of condensed matter physics with the possible early-Universe phase transitions and associated gravitational waves. Of crucial importance for the spectrum of generated gravitational waves are the lifetime of the metastable phase and the propagation speed of the expanding phase boundary during the transition [5, 6]. The simplest model, the homogeneous nucleation via thermal fluctuations by Cahn and Hilliard [7] and Langer [8], predicts a lifetime of the metastable supercooled $^3$He-A longer than the age of the Universe under relevant experimental conditions [9, 10]. The required large critical radius of the B-phase bubble ($R_c = 2\sigma_{AB}/\Delta F_{AB} \sim 1$ μm), at which the gain in bulk free energy density $\Delta F_{AB}$ overcomes the surface tension $\sigma_{AB}$ and leads to an expanding bubble, is the underlying reason for this. An alternative intrinsic mechanism for $^3$He-B to grow past the critical size is quantum tunnelling, for which the theory also suggests an extremely slow nucleation rate [10, 11]. Resonant tunnelling effects, involving intermediate phases between $^3$He-A and $^3$He-B, have been proposed as a potential pathway to decrease the lifetime of supercooled A phase at certain values of temperature $T$, pressure $P$, and magnetic field strength $B_0$ [12].

Experimentally the phase transition between the A and B phases, usually from supercooled $^3$He-A into $^3$He-B, is routinely observed. This has led to proposals and studies of various heterogeneous nucleation mechanisms. Textural surface singularities (boojums) by themselves have been calculated to be unlikely to encourage the nucleation enough [9, 10]. Surface roughness, in particular where the sample container does not have specifically smoothened surfaces or is not isolated from a sintered heat exchanger, has been found to result in a cooling-rate dependence [13], a nucleation location dependence [14], or even a cooling-path-dependence [15, 16] of the A-to-B nucleation temperature. This has been sometimes accompanied by a catastrophe line beyond which the primary supercooling of $^3$He-A is not possible [17, 18] and by a history dependence of the degree of supercooling during the secondary nucleations [19].

It has been proposed that in the case of secondary nucleations, tiny surface pockets can hold the stable phase by surface tension, and thus retain a memory of it, eventually expanding and filling the sample volume (the "lobster-pot" scenario) [9, 20]. Similarly, stabilisation of seeds or precursors of the B phase near sharp surface defects and inside surface cavities has been discussed in the context of primary nucleations [15, 16, 21]. In most of the experiments referred above, the transitions were only observed on cooling; thus the interplay between surface roughness, textural singularities, and the superfluid flow due to thermal gradients cannot be ruled out. In the presence of rough surfaces, mechanical vibrations of the cryostat have also been observed to trigger the nucleation [20, 22].

The third category is a phase transition triggered by ionising radiation working alone or in combination with the above-mentioned surface and textural effects.

This radiation can stem either from cosmic rays or from radioactive materials near the experiment. Two models have been proposed to explain such phase transitions: baked Alaska mechanism [23, 24] and cosmological (Kibble–Zurek) scenario [25, 26]. The models differ in the treatment of the dissipation of heat within the superfluid and in the way that the transition to superfluid phases takes place during cooling [27–29]. However, at the heart of both scenarios is the local deposition of energy by a passage of a particle through supercooled $^3$He-A. Subsequent heating of a tiny region of $^3$He above the superfluid transition temperature $T_c$ is followed by a rapid cooling, or quench, during which an emergence of a single phase [29] or multitude of random order-parameter domains [30] takes place. If at any point during the process a B-phase bubble larger than the critical size comes into existence, the B phase will take over and eventually expand to fill the whole sample volume.

The radiation-triggered phase transition is a stochastic process. This is firstly due to the random nature of particle interactions and the distribution of deposited energy within the sample [24] and secondly due to the random nature of the creation of a B-phase bubble within the sample following the quench. In order to study this process as independently of the other mechanisms as possible, one requires a sample in thermal equilibrium and a smooth-walled sample container isolated from the rough surfaces in the system, especially from the heat exchanger.

Such experiments, performed near melting pressure at 34.2 bar, have studied the lifetime of the supercooled metastable A phase as a function of incoming particle flux, temperature, and magnetic field strength [24, 31, 32]. The measured mean lifetime, $\tau$, is compared to the baked Alaska model, the latest version of which predicts $\tau = C_0 \exp\left([R_c(T,P,B_0)/R_0(T,P)]^N\right)$. Here constant $C_0$ depends on the radiation type and activity and $N = 3$. The characteristic length $R_0$ should be fitted to the data and is expected to vary substantially with pressure but only weakly with temperature at low temperatures where the lifetime of the metastable A phase usually is measured [24, 29]. The critical B-phase bubble radius $R_c$ was originally given in Ref. [9] as the Cahn-Hilliard radius, with a range $N = 3 - 5$ in the exponent, but in Ref. [24] a more exact $R_c$ has been derived together with a single-valued exponent, $N = 3$. The experiments at 34.2 bar found the best fit with $N = 3/2$ [32].

Our target is to access possible intrinsic phase-nucleation mechanisms. One first needs to understand and/or eliminate all the external triggers which may dominate. Here we present a versatile nanofabricated stepped-height sample container for SQUID-based nuclear magnetic resonance (NMR) studies. In such a container, the state of the superfluid $^3$He is tuned by confinement, which allows creating several essentially isolated volumes of superfluid within a single measurement platform [33, 34]. Our container has nearly atomically smooth surfaces in five such "isolated" phase-nucleation volumes, eliminating the heterogeneous nucleation. The sample volumes are orders of magnitude smaller than in the earlier experiments, significantly decreasing the event rate from cosmic rays and radioactivity in the materials surrounding them. We observe very strong supercooling of $^3$He-A, superheating of $^3$He-B, and stochastic transitions in both directions, with no evidence for a cooling/warming-rate dependence or catastrophe lines. This will allow us to measure the lifetime of supercooled A phase over a wide temperature and pressure range in the future.

## 2 Experimental Setup

### 2.1 Design of Step-Height Nanofluidic Sample Container

The nanofluidic sample container for SQUID-based NMR studies (Fig. 1a) has a confinement volume with stepped cavity height $D$. The sample space between two silicon wafers consists of five $D = 6.8$ μm phase-transition volumes ("lakes") and a strongly-confining $D = 75$ nm isolation barrier ("shore") surrounding them (Fig. 1b). The depth of the lakes is such that they contain essentially bulk superfluid: $D \gg \xi_0$, where $\xi_0$ is the Cooper-pair diameter tuneable by pressure between

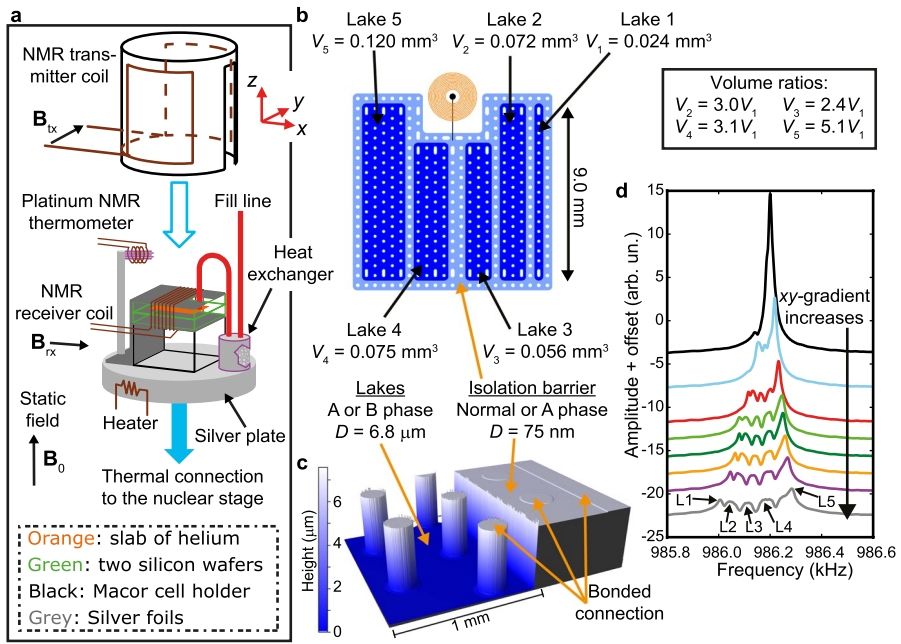

**Fig. 1** **a** The experimental platform for SQUID NMR. **b** Map of the nanofluidic sample container, showing five *dark blue* phase-nucleation volumes (lakes) surrounded by shallow *light blue* isolation barrier (shore). The support pillars are shown as *white dots*. The concentric *orange circles* are the contact grooves around the fill-line hole (*black circle*). The deep trench providing entry from the fill line into the shore region is shown as *black line*. **c** An optical profilometry scan of a fabricated sample container. A blank lid wafer is fusion bonded to the patterned wafer along the edge of the cavity and on top of each support pillar. The spikes around the edges of height steps are imaging artefacts. **d** NMR spectra as a function of strength of magnetic field gradient in the $xy$-plane across the lakes at 0 bar at $T > T_c$. The lowest spectrum shows the chosen configuration for the experiments presented in this work, optimising between the relative separation and the signal line width arising from different lakes (identified as L1, L2, L3, L4, and L5). The lines are shifted vertically for clarity

20 and 80 nm. Thus, in the lakes the order parameter is suppressed from the bulk value only within a few $\xi_0$ from the surfaces [35], with a minimal effect on the phase diagram. Each of the five lakes has a different surface area, and hence volume, to potentially provide more insight into the nucleation mechanisms. The surrounding shallow region functions as a confinement-based isolation barrier between the lakes and the rest of the system, such as the rough surfaces of the heat exchanger. Throughout the 75 nm high shore, the order parameter is affected by the anisotropic pair breaking at the surfaces. Here at any pressure the effective cavity height $D/\xi_0 < 5$, which entirely suppresses the isotropic B phase. Only normal liquid $^3$He or the anisotropic superfluid $^3$He-A can exist in such a confinement, with both $T_c$ and the order-parameter amplitude being tuneable by the quasiparticle scattering boundary condition [33, 35].

The fill line entry hole and the confining cavity are connected by a 2.25 mm long, 20 μm wide, and 110 μm deep trench providing good thermal connection and minimising the possibility of a cavity collapse due to differential thermal contraction between the silicon cell and the silver fill line (Fig. 1b). For the same purpose, a custom laser-machined silicon washer is attached with epoxy (Stycast 1266 mixed with silicon powder) around the entry hole between the fill line and the wafer, as in the previous generation of nanofluidic cells [33].

In the confinement volume, there are 150 μm wide support pillars arranged in a hexagonal pattern 500 μm apart from each other (Figs. 1b, c) to reduce the cavity height distortion by pressure. From finite-element modelling, we estimate a distortion with pressure of the order 1 nm/bar. To accommodate the lakes of regular shape and to keep the distortion at minimum, pillar density around the cavity edges is increased and some pillars are elongated to oval shapes.

## 2.2 SQUID NMR

The experimental platform, based on the SQUID-NMR spectrometer [36] used in our earlier work [33, 37–39], is shown in Fig. 1a. The silicon sample container (cell) sits on a Macor holder anchored on a silver plate which is thermally weakly linked to the copper nuclear demagnetisation stage of the cryostat. This way we can independently control the temperature of the silver plate by powering a heater attached to it up or down. Silver foils (25 μm thick) are glued with conductive silver paint on the top and bottom surfaces of the cell and clamped down on the silver plate to cool the silicon which is coated with a 500 nm thick layer of evaporated silver. A silver fill line connects the entry hole in the cell to the 13 m$^2$ sintered silver heat exchanger on the silver plate.

Nuclear spin precession is excited by a saddle-shaped transmitter coil on a Kel-F former around the cell holder. We use small NMR pulses (spin-tipping angle $\beta < 1°$) to virtually remove the dependence of the superfluid precession frequency on the pulse size. The pulses are applied once every three seconds, providing a good signal quality after 60 averages. The NMR receiver coil is part of a series tuned circuit having a resonance frequency $f_0 = 965$ kHz and a quality factor $Q = 30$. The tiny signal is amplified with a two-stage SQUID device [40] operated in flux-locked loop mode, monitored and controlled by room-temperature Magnicon XXF-1 electronics [41].

Since the gyromagnetic ratio of $^3$He is $\gamma/(2\pi) = 32.43$ MHz/T, our NMR scheme requires a static magnetic field strength $B_0 \approx 30$ mT. The field is oriented perpendicular to the cavity surfaces, $\mathbf{B}_0 = B_0\hat{\mathbf{z}}$. The field strength is fine-tuned to locate the $^3$He Larmor frequency away from any prominent noise features in our system: $f_L = \gamma B_0/(2\pi) = 986$ kHz. Additionally, a small magnetic field gradient is applied in the $xy$-plane to separate the precession frequencies of the five lakes from each other to allow independent monitoring of each of them (Fig. 1d).

The temperature of the heat exchanger on the silver plate, $T_{sp}$, is measured with a Pt-NMR thermometer calibrated against a $^3$He melting-curve thermometer. This temperature can be precisely controlled with the heater shown in Fig. 1a.

The NMR precession frequency $f_L$ of normal liquid $^3$He is defined by the magnetic field alone. In the superfluid phases the precession frequency $f$ shifts by $\Delta f = f - f_L$, which can be either positive or negative, depending on the curvature of the dipole energy as a function of rotations in the spin space [42]. In our sample geometry $\mathbf{B}_0 \parallel \hat{\mathbf{z}}$ is normal to the cavity surfaces, forcing $\hat{\mathbf{d}} \perp \hat{\mathbf{z}}$ in $^3$He-A. The order-parameter vector $\hat{\mathbf{d}}$ points in the direction of zero spin projection. The lakes are deep enough not to suppress the overall order-parameter amplitude ($D \gg \xi_0$) but shallow enough to lock the orbital angular momentum of all the A-phase Cooper pairs, $\hat{\mathbf{l}} \parallel \hat{\mathbf{z}}$, with no textural deformation present in our relatively high magnetic field [43]. The relevant length scale for the latter is the dipole length $\xi_D \sim 10\mu$m [42]. This dipole-unlocked configuration maximises the dipole energy, resulting in $\Delta f = -\Delta f_A$, i.e. a frequency shift with the same magnitude and opposite sign to the bulk A-phase frequency shift $\Delta f_A > 0$ [33, 37]. In $^3$He-B the dipole energy minimum (spin-orbit rotation about axis $\hat{\mathbf{n}} = \hat{\mathbf{z}}$ by angle $\theta = 104°$) is compatible with our geometry. This has a small positive frequency shift due to the magnetic field [42] but again no textural effects are expected. Since $D < \xi_D$, the metastable orientation of the B phase with $\hat{\mathbf{n}} \perp \hat{\mathbf{z}}$ and $\theta = \pi$ may also stabilise in the lakes, characterised by $\Delta f < 0$ [38].

### 2.3 Nanofabrication

The nanofabrication techniques have been modified compared to those used in the earlier work [44, 45]. The sample region is patterned using contact lithography and etched directly into a silicon wafer, on top of which a blank silicon wafer is fusion bonded. Instead of the repetitive growth and etching of thermal silicon dioxide, the cavity region in the container is defined by a combination of hydrogen-bromide-based reactive ion etching (HBr RIE) and deep reactive ion etching (DRIE). The benefits of this approach are the speed and simplicity of the process, the possibility for smaller and more complicated features, and much steeper sidewalls. The latter is especially relevant in the case of deep features used here. The other major deviation from the earlier fabrication recipe is the growth of thermal SiO$_2$ on all the surfaces before fusion bonding the wafers (see below). We confirmed by AFM (Bruker Icon with a sharp TESP-SS tip having a spring constant of 42 N/m and a radius of curvature of $3.5 \pm 1.5$ nm used in tapping mode) that the HBr-based RIE process results in rms surface roughness of 0.6 nm after a 200 nm deep etch and the subsequent thermal oxide growth. This is the same value we found outside the etched region. The DRIE process has been shown to result in extremely smooth surfaces with even smaller 0.2 nm rms roughness after a 10 μm deep etch [46]. Such surfaces are somewhat rougher than the atomically smooth surfaces created by the oxidation method [45] but should easily prevent the trapping of seeds of distorted order parameter of size $\sim \xi_0$.

The step-by-step nanofabrication process flow is illustrated in Fig. 2 and described here in detail. The fabrication is done on a 525 μm thick 4-inch-diameter silicon wafer. Altogether 12 sample containers of two designs (the 5-lake variation used here together with a simpler 1-lake variation) are patterned and fabricated using the following process. The 5-lake variation was chosen to be studied first to allow the characterisation of the volume effect on the phase transitions and a more efficient collection of statistics in the case of a stochastic process dominating the experimental observations.

(a) Shallow $D = 75$ nm rectangular cavities are etched over the confinement regions using a single RIE step.
(b) A series of DRIE steps, 18 in total, alternating between passivation and etching, is used to create the 6.8 μm deep lakes within each cavity: this results in an almost vertical sidewall angle of 89.6°, which is strikingly different to the tapered sidewall created by the oxidation method [45]. Each step in the process creates a smooth sidewall scallop with height of 0.37 μm and maximum depth of 0.05 μm on average, as inferred from SEM images of a test structure etched using the same DRIE device and recipe. The vertical sidewall ensures that the support pillars within the lakes maintain constant diameter over the full depth

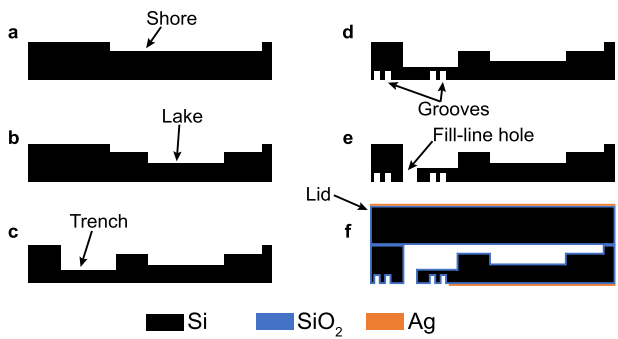

**Fig. 2** The nanofabrication process flow. Each layer is defined with photolithography, using either photoresist SPR220-3.0 or AZ 12XT as a protective layer during the etching. **a** The overall cavity region is etched with HBr RIE down to 75 nm depth. The 6.8 μm deep lake regions (**b**), the 110 μm deep thermal-link trench (**c**), the concentric 20 μm deep grooves (**d**), and the through-hole for the helium to enter the cavity from the fill line (**e**) are etched with DRIE. **f** The wafer is passivated with a 50 nm thick thermal oxide before fusion bonding together with an identically cleaned and passivated blank wafer (lid). After dicing, a 500 nm thick layer of Ag is evaporated onto the front and back of the individual sample container to provide a better thermal link to Si. More details are described in the text. The diagram is not drawn to scale

and that there will be no confinement-defined phase boundary between the A and B phases along the sidewall [47]. Instead, each lake is expected to be completely filled by either $^3$He-A or $^3$He-B at any temperature $T < T_c$.

(c) Within each cavity, a single 110 μm deep trench is etched using DRIE, connecting the confinement volume to the fill line. In total 265 steps are required, resulting in an equally large number of sidewall scallops. The trench must not reach into any of the lakes to keep them pristine for phase-nucleation studies.

(d) Overall, 13 concentric 20 μm deep and 20 μm wide circular grooves (740–2400 μm in diameter) are DRIE etched on the back side of each of the sample container on the wafer. These grooves provide surface contact for glueing down the silicon washer leading to the helium fill line [45].

(e) A 300 μm diameter entry hole is etched through the wafer from the back side up at each centre point of the circular grooves, using DRIE. The process wafer is mounted on an oxide-coated carrier wafer to provide a cavity-side protection during etching. Here the recipe is different and designed for etching extremely high aspect ratio features with a small open area. The resulting scallops on the 88.4° sidewalls of the hole are 3 μm high and 0.7 μm deep. The hole overlaps with the 110 μm deep trench, completing the path between the fill line and the cavity (see Fig. 1b for details).

(f) After all the features are patterned, the cells are characterised with an optical profilometer (Zygo NewView 5000), which creates a 3D image of part of the cell (Fig. 1c). The series of such images confirms the dimensions of the fabricated features, most importantly the etched depths of the lakes and the shore, and their spatial uniformity. The height variation within a single cell was measured to be less than 10 nm.

(g) The wafers are thoroughly cleaned in solvent followed by a two-step RCA cleaning process (SC1 → HF → SC2). Immediately after cleaning, as a last step before bonding, a dry 50 nm thick thermal $SiO_2$ is grown on all the surfaces inside a CMOS clean furnace. The purpose of the silicon dioxide layer is chemical passivation [48, 49]. This is essential to minimise the effects of high charge density or dangling bonds on the superfluid within the cavity; in previous work, in the absence of this passivation, it was found to be impossible to grow a uniform thin superfluid $^4$He film as a preplating layer on the cavity surface.

(h) Finally, the patterned wafer is fusion bonded to a similarly cleaned and passivated blank wafer in vacuum, using a SUSS MicroTec SB6e wafer bonder. We ensure, with retractable spacers, that the cavities are fully pumped out before contact is made. The bonding strength and quality are increased by annealing in $N_2$ atmosphere at 1100°C. The resulting bond quality is assessed with an infrared microscope, followed by dicing to separate the individual sample containers from each other.

(i) Before connecting the silicon washer and silver fill line around the entry hole, a 500 nm thick layer of silver is thermally evaporated onto the top and bottom sides of the container. This improves the thermal link between the silicon and the silver foils glued on it. During Ag evaporation, a physical mask is used to avoid depositing any silver on the outer sidewalls (which would prevent the NMR excitation and detection) or into the fill-line hole (which would compromise the purity of the cavity).

## 3 Results

### 3.1 Constant-Rate Temperature Sweeps

To test the performance of our experimental setup, we study the superfluid phase transitions at pressure $P = 22.0$ bar, above the polycritical point, to allow a finite temperature range of stable $^3$He-A just below $T_c$. At this pressure and $B_0 \approx 30$ mT, the bulk phase-transition temperature between $^3$He-A and $^3$He-B is $T_{AB} \approx 0.96 T_c$ [50], which corresponds to 2.20 mK on the Greywall temperature scale [51]. We measure the level of supercooling achievable in our setup by ramping the silver plate temperature down at a constant rate from normal state through $T_c$ until all the lakes have transitioned from $^3$He-A into $^3$He-B. The transition is detected by observing a jump in the NMR precession frequency.

Two such examples at slow cooldown rates are shown in Fig. 3 (with details highlighted in Fig. 4). Two important notions become immediately clear here: (1) The lowest temperature of supercooled $^3$He-A in these examples is $T = 1.35$ mK $= 0.6 T_c$, lower than observed before at the same pressure [32]. (2) The phase transitions in the lakes do not take place at the same temperature each time—indicating a lack of catastrophe line— as expected for the atomically smooth sample container. We believe the phase transitions triggered during the cooling are not related to the cooldown rate itself, but rather are due to the stochastic $T$-dependent mechanism discussed in the next section.

We determine the helium temperatures $T_{He}$ inside the individual lakes, when in the A phase, from their NMR precession frequency shifts $\Delta f$, using the known temperature dependence of the bulk A-phase frequency shift $\Delta f_A$ determined inside a different nanofluidic cell with thin cavity, $D = 80$ nm [52], and specular quasiparticle scattering boundary condition reproducing the bulk-like behaviour [33]. In this work we study pure $^3$He with no added $^4$He to cover the surfaces, and thus have strongly pair-breaking solid $^3$He boundaries. However, in the $D = 6.8$ μm $\gg \xi_0$ lakes the order parameter suppression induced by any boundary condition is negligible and can be omitted in the analysis here, giving $\Delta f = -\Delta f_A$. There is a significant temperature gradient between the silver-plate temperature $T_{sp}$ and the helium temperature $T_{He}$, shown in Fig. 3c, stemming from an unusually high thermal impedance between the confinement volume and the heat exchanger. We attribute this to a partially collapsed nanofabricated trench or a partially blocked fill line connecting these two volumes. This at present limits our ability to cool down the sample rapidly and to reach the lowest temperatures, but nevertheless allows us to measure the lifetime of supercooled $^3$He-A under various static conditions over a reasonably wide temperature and pressure range. Additionally, there is a much smaller (less than 20 μK) temperature gradient across the confinement volume as seen in Fig. 4a: Lake 4— which is closest to the fill-line connection trench—is the coldest, whereas Lake 1— the longest distance away—is the hottest (see Fig. 1b for the geometry). Although this results in a small time delay between the superfluid transitions in different lakes, for clarity in Figs. 3 and 6 only the times of Lake 4 crossing $T_c$ are indicated.

The A-to-B phase transition is associated with latent heat release [53] which can be seen as sudden precession-frequency jumps in the lakes remaining in the

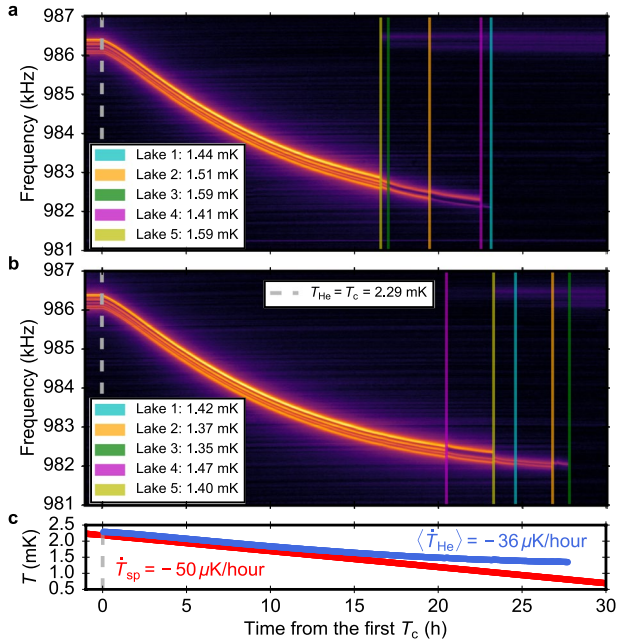

**Fig. 3 a, b** The colour-coded NMR spectra on two cooldowns at $P = 22.0$ bar. The NMR precession frequencies in all five lakes are monitored simultaneously. The *dark colour* indicates the noise background whereas the five brighter *yellow-red* traces moving from 986 kHz down with time are the NMR signatures arising from the lakes. The lowest-frequency signal arises from the smallest lake (Lake 1) and the highest-frequency signal from the largest lake (Lake 5) with all the other lake volumes in between (see Figs. 1d and 4). Below $T_c$ (*vertical dashed line*) all the lakes show a negative frequency shift and unchanging signal amplitude, i.e. sample magnetisation, as is expected for the equal-spin-pairing $^3$He-A. Eventually (between 15 and 30 hours after passing through $T_c$ in these examples) the signals disappear one by one at the times indicated by *coloured vertical lines*. At these same times low-amplitude features appear close to the corresponding Larmor frequencies of different lakes as NMR signatures of low-temperature $^3$He-B. Thus the marked times, and extracted helium temperatures shown in the legend boxes at those times, indicate phase transitions from $^3$He-A to $^3$He-B within the lakes. **c** Temperature of the silver-plate (*red line*), $T_{sp}$, during the two measurement runs. The steady sweep rate is controlled by powering down the heater attached to the plate. *Blue line* shows the helium temperature determined from the A-phase frequency shift of Lake 3 in panel **b**. The sweep rate slows down as the silver-plate temperature decreases, so only the average value is given

A phase after each transition event (Fig. 4b). The bigger the lake which transitions the larger the amount of heat released: phase transitions in Lake 5 heat the other lakes the most. The increase in temperature (up to 20 µK) depends on the level of supercooling in the transitioning lake and the total heat capacity of the superfluid in all the lakes combined.

### 3.2 Stochastic Phase Transitions

To study the lifetime of the supercooled A phase at constant temperature, we first stabilise $T_{sp}$, after which a suitably chosen NMR heating pulse is applied to reset the experiment. This pulse heats all the helium in the lakes above $T_c$ via the parasitic local heating within the nanofluidic cavity [39]. Such pulses have a minimal effect on $T_{sp}$. Moreover, we note here that the $^3$He in the lakes is not significantly heated by the small NMR probing pulses ($\beta < 1°$) applied every three seconds to monitor the state of the superfluid phase within them.

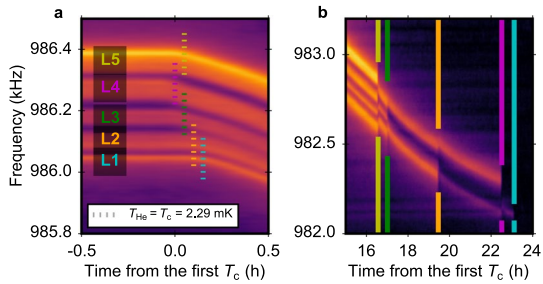

**Fig. 4** Highlight of the details of the cooldown shown in Fig. 3a. **a** Above $T_c$ all five lakes are clearly visible and separated as high peak features in the NMR spectrum. They are however neither fully independent nor Lorentzian peaks due to the field inhomogeneity and spectral overlap (see also Fig. 1d). During a slow cooldown, the small temperature gradient across the confinement volume results in different times of crossing $T_c$ in each lake (*coloured dotted lines*): Lake 4, the coldest one, transitions first; Lake 1, the hottest one, transitions last. Due to this gradient, Lake 1 and 2 peaks merge to form a single composite peak less than 200 μK below $T_c$. **b** During each transition from $^3$He-A into $^3$He-B (*coloured vertical lines*), one of the peaks in the NMR spectrum in the A-phase region disappears (or in the case of composite peaks, the peak amplitude suddenly drops). The surrounding lakes are also affected, since the loss of one of the partially overlapping peaks decreases the spectral weight of the nearby peaks as well, lowering their signal intensity. Frequency-shift jumps in the lakes remaining in the A phase after the transitions are due to the associated heat release

After the heating, we wait until the helium cools down and all the lakes have transitioned from the A phase to the B phase. Two such examples at $P = 22.0$ bar and $T_{sp} \approx 1.045$ mK are shown in Fig. 5. The average cooling rates here are an order of magnitude higher than in the previous section ($\dot{T}_{He} \approx -500$μK/h). We again do not observe any high-temperature phase transitions, and the helium reaches a stable temperature before the transitions take place. Thus, the system seems to be independent of heterogeneous cooling-rate dependent nucleation mechanisms and of a high-temperature catastrophe line. Instead, a stochastic lifetime of supercooled $^3$He-A is observed, varying strongly between the lakes and between the cooldowns. Similar behaviour is reported in Refs. [24, 32] where the lifetimes from individual measurements are shown to follow an exponential distribution. This has been attributed to local heating events and subsequent quenches through $T_c$ caused by cosmic rays, setting in motion random occurrences of B-phase nucleation. The radiation sources in the laboratory have been shown to accelerate the nucleation rate. Following the approach of Refs. [24, 32], we aim to use our platform to collect nucleation statistics to determine the temperature and pressure dependence of the mean lifetime $\tau$ of the supercooled A phase and to compare it to the predictions of different theoretical models, such as the baked Alaska model [29].

Over the course of these experiments, we also confirmed that the B-phase nucleation is not triggered by mechanically vibrating the cryostat; neither the gentle

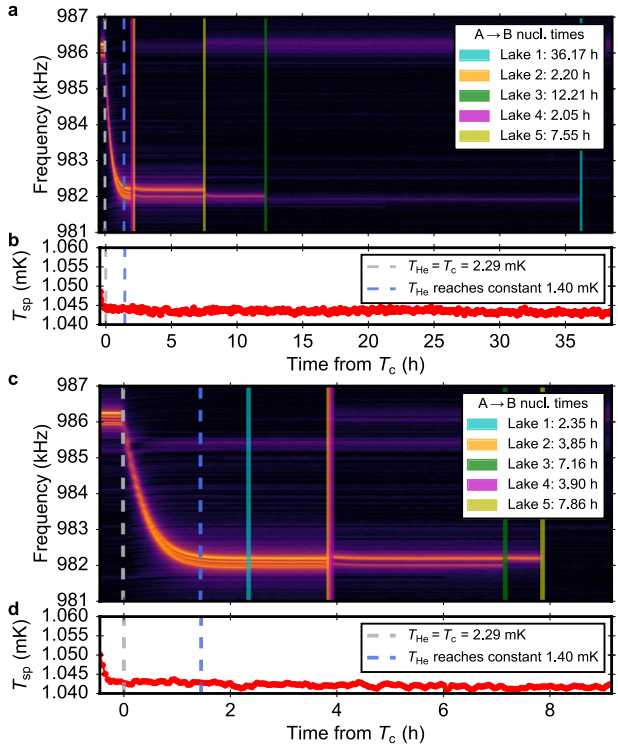

**Fig. 5 a, c** Evolution of the NMR precession frequencies of the five lakes after a high-amplitude NMR pulse heats all the helium into the normal state. Helium cools down below $T_c$ at time zero (*grey dashed line*) and reaches constant temperature, $T_{He} = 1.40$ mK, less than two hours later (*blue dashed line*). During the fast cooldowns all the lakes cross $T_c$ within one time step (three minutes). The $^3$He-A to $^3$He-B transition times within each lake volume are marked as *coloured vertical lines* and listed inside the legend box. **b, d** The silver-plate temperature, $T_{sp}$, is kept constant during these measurements by a heater. The scales of the time axes of these two example measurements are different but the cooldown rates are similar tapping nor the more vigorous vibrations associated with transfer of liquid $^4$He coolant into the dewar surrounding the cryostat resulted in phase transitions, unlike in the earlier experiments with rough surfaces or impurities present in the sample volume [13, 20, 22].

### 3.3 Superheating

A first-order phase transition can manifest superheating as well as supercooling, and both have previously been studied in superfluid $^3$He [17, 19, 54, 55]. However, in those experiments, as well as in analogous studies where the phase transitions were driven by a magnetic field [20], a history dependence has been detected. A possible reason for this is the proposed "lobster-pot" scenario causing secondary nucleations of $^3$He-A [9]. In order to avoid that, and to observe an irreversible superheated transition from $^3$He-B to $^3$He-A, one must prevent both the existence of trapped seeds of $^3$He-A on the surfaces and the stabilisation of $^3$He-A in regions close to the corners of the sample container, where it may survive even when $T \ll T_{AB}$ [47]. The extremely smooth surfaces and vertical walls of the lake volumes in our sample container achieve this by design.

The shallow shore region would also act as a nucleation source if it was filled with the A phase. Here this possibility is eliminated by choosing the shore depth $D = 75$nm, pressure $P = 22$bar, and strongly pair-breaking solid $^3$He surfaces. This completely suppresses the superfluidity in the shore down to approximately 1.70 mK [33], well below bulk $T_{AB} \approx 2.20$mK.

On a slow warm up from below $T_{AB}$, shown in Fig. 6, all the lakes which originally contained $^3$He-B are superheated above $T_{AB}$. In one lake the B phase survives to within 10 μK from $T_c$. Upon repetition of the warm-up experiment, the superheating is always detected. The observed B-to-A transition temperatures suggest a stochastic process may be also responsible for the nucleation of $^3$He-A out of $^3$He-B.

We must consider secondary nucleations, since the A phase existed in the lakes while cooling the sample down from above $T_c$ prior to starting the warm-up experiment. However, the smooth surfaces, regular container geometry, and normal $^3$He in the shore make the survival of the A-phase seeds after the A-to-B transition in each lake highly unlikely. Thus, we argue that primary nucleations not only of the B phase but also of the A phase are likely to be observed in our experiments. This opens up the possibility for studying the mechanism of the phase transition between these two phases in both directions.

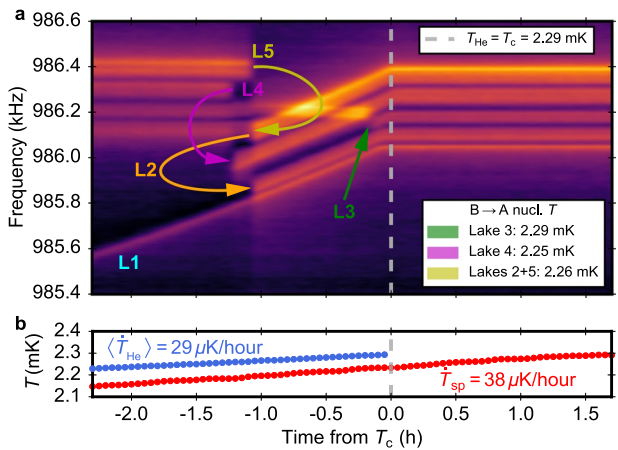

**Fig. 6** **a** NMR precession frequency traces arising from the five lake volumes during warm up. Lakes 2–5 transition from the B phase precessing near $f_L$ into the A phase with negative $\Delta f$ at the times indicated by *coloured arrows*. Each arrow points at the first appearance of the A-phase peak for the lake in question. Transitions in Lakes 2 and 5 take place within the same three-minute time step of the experiments. The B-phase peak of Lake 3 overlaps with the A-phase peaks of Lakes 4 and 5 as $T$ increases, distorting the precise tracking of the individual peaks during the overlap. The transition temperatures, all of which indicate superheating of $^3$He-B above $T_{AB} \approx 2.20$mK, are extracted from the A-phase frequency shifts immediately after the transitions. Lake 1 has $^3$He-A in it throughout the warm up. **b** Temperature of the silver plate (*red dots*), controlled with the heater, is swept up through $T_{AB}$ and $T_c$ (*grey dashed line*) at a constant rate. *Blue dots* show helium temperature in Lake 1 below $T_c$. Time axis zero is locked at the time of crossing $T_c$ in Lake 4

We can tune the stability of the A phase in the shore with pressure and/or $^4$He preplating [33]. This will allow the operation of the shore as a nucleation "seed" of precisely defined geometry and the stimulation of secondary nucleations of the A phase.

## 4 Discussion and Conclusion

Our preliminary results show that nanofluidic samples with stepped confinement are a promising tool to investigate the first-order phase transition between $^3$He-A and $^3$He-B. The approach relies on the creation of isolated tiny "lakes" of superfluid, which can be cooled through a surrounding very thin "shore" region of either $^3$He-A or normal fluid. We determine the superfluid phase in each lake by sensitive SQUID NMR combined with simple imaging techniques. With nearly atomically smooth surfaces and vertical walls, the phase-transition volumes allow both very strong supercooling of $^3$He-A and superheating of $^3$He-B, so far confirmed at a single pressure, $P = 22.0$ bar. This work will be extended over a wide pressure and temperature range, including below polycritical point. Of particular interest is the low-pressure regime where very little supercooling has been observed under stronger confinement, potentially linked to resonant tunnelling [47]. We anticipate that data collection will be significantly accelerated by combining measurements from all five lakes.

The geometry and the surface quality of the nanofluidic sample container are designed to eliminate seeds triggering heterogeneous nucleation; confirmation of a lack of history dependence is a future task. It remains likely that the detected stochasticity has a contribution from external triggers, such as cosmic rays and radioactivity in the laboratory environment. The effect of both of these can be minimised by additional shielding or careful choice of materials. We will characterise the radioactivity of the specific materials present in significant amount close to the sample, including Macor, silicon, and Kel-F. Combining this information with the recent assay result of other commonly used materials in the cryostat [56, 57], the modelling of corresponding energy deposition rate and distribution in the sample volumes becomes possible.

Eventually the understanding of the extrinsic effects, coupled to a study of nucleation across the entire $P - T$ plane of the superfluid phase diagram, should uncover potential underlying intrinsic mechanisms and their cosmological analogues. Besides the lifetime of the metastable phase, the kinetics of the phase transition—in particular the speed of the propagating phase boundary—are crucial when comparing the A-B phase transition to the cosmological phase transitions and the predicted spectrum of generated gravitational waves [5, 6].

Furthermore, the successful design, fabrication, and use of a nanofabricated stepped-height sample container is the latest step into the microkelvin studies of topological mesoscopic superfluidity. The method enables the creation of hybrid nanofluidic devices by sculpture of the order parameter [34]. Future variations of such structures—combining the flexibility of confinement geometry with *in situ* tuneable boundary condition [33]—provide unprecedented versatility to study and harness the emergent topologically protected surface-, interface-, and edge-bound low-energy states [34, 58–61].

**Acknowledgements** We thank J. A. Sauls and G. Cowan for the helpful discussions. The sample container used in this work was nanofabricated at the University of Michigan Lurie Nanofabrication Facility. Measurements were made at the London Low Temperature Laboratory, where we acknowledge the excellent support of technical staff, in particular Richard Elsom, Ian Higgs, Paul Bamford, and Harpal Sandhu.

**Author Contributions** The experiments and data analysis were carried out by P.J.H. with contributions from L.V.L. The nanofluidic sample container was fabricated, characterised, and prepared for experiments by N.E. with contributions from A.C., A.S., and X.R. The manuscript was written by P.J.H. with contributions from all authors. The experimental program was devised by P.J.H., L.V.L., J.M.P., and J.S. The project was supervised by J.M.P., A.C., and J.S.

**Funding** The research leading to these results has received funding in the UK from the UKRI STFC under ST/T006749/1; ST/T006773/1; ST/T00682X/1 ("QUEST-DMC") and from the UKRI EPSRC under EP/R04533X/1 and EP/W015730/1, and at Cornell from the NSF under DMR-2002692. In addition, the work has been supported by the European Union's Horizon 2020 Research and Innovation Programme under Grant Agreement no 824109 (European Microkelvin Platform).

**Data Availability** The authors declare that the data supporting the findings of this study are present within the paper.

## Declarations

**Conflict of interest** The authors declare no Conflict of interest.

## Authors and Affiliations

**Petri J. Heikkinen[1] · Nathan Eng[1] · Lev V. Levitin[1] · Xavier Rojas[1] · Angadjit Singh[1,2] · Samuli Autti[3] · Richard P. Haley[3] · Mark Hindmarsh[4,5] · Dmitry E. Zmeev[3] · Jeevak M. Parpia[6] · Andrew Casey[1] · John Saunders[1]**

✉ Petri J. Heikkinen
  petri.heikkinen@rhul.ac.uk

1. Department of Physics, Royal Holloway, University of London, Egham TW20 0EX, UK
2. Clarendon Laboratory, Department of Physics, University of Oxford, Oxford OX1 3PU, UK
3. Physics Department, Lancaster University, Lancaster LA1 4YB, UK
4. Department of Physics and Astronomy, University of Sussex, Falmer, Brighton BN1 9QH, UK
5. Department of Physics and Helsinki Institute of Physics, University of Helsinki, FI-00014 Helsinki, Finland
6. Department of Physics, Cornell University, Ithaca, NY 14853, USA